\documentclass[twocolumn]{revtex4}[12pt]
\usepackage{graphicx}
\begin{document}
\title{Solitary potentials in a three component space plasma}
\author{M. Sarker$^{a*}$, B. Hosen$^{a}$, M. R. Hossen$^{b}$, and A. A. Mamun$^{a}$}
\address{$^{a}$ Department of Physics, Jahangirnagar University, Savar,
Dhaka-1342, Bangladesh. $^{b}$ Department of General Educational
Development, Daffodil International
University, Dhanmondi, Dhaka-1207, Bangladesh\\
 $^*$Email: sarker.plasma@gmail.com}
\begin{abstract}
The heavy ion acoustic solitary waves (HIASWs) in a magnetized,
collisionless, three component space plasma system in the presence
of dynamical heavy particles and bi-kappa distributed two distinct
temperature of electrons has been carried out. The Korteweg-de
Vries (K-dV), the modified K-dV (mK-dV), and the further mK-dV
(fmK-dV) equations are derived by employing the well known
reductive perturbation method. The basic features of HIASWs
(speed, amplitude, width etc.) are found to be significantly
modified by the effects of number density of plasma species,
superthermal electrons with two distinct temperatures and the
magnetic field (obliqueness). The K-dV and fm-KdV solitons exhibit
both compressive and rarefactive structures, whereas the mK-dV
solitons support only compressive structures. In order to trace
the vicinity of the critical heavy ion density, neither the
magnetized KdV nor magnetized mKdV equation is suitable for
describing the HIASWs. Therefore, we have considered a higher
order nonlinearity to describe the solitary profiles by deriving
fmK-dV equation. The implication of our
results in some space and laboratory plasma situations are concisely discussed.\\

\end{abstract}

\maketitle
\section{Introduction}
%%%%%%%%%%%%%%%%%%%%%%%%%%%%%%%%%%%%%%%%%%%%%%%%%%%%%%%%%%%%%%%%%%
Recently, propagation of linear and nonlinear waves in
electron-ion (EI) plasma has generated a lot of interest among
the researchers
\cite{Chen1984,Durrani1979,Davidson1972,Chawla2013,Mamun1997}.
Since many nonlinear fascinating structures like solitary waves,
solitons, shocks, double layers and so on are observed in space,
astrophysical, and laboratory plasmas, so a large number of
investigations are going on it
\cite{Hossen2014a,Hossen2014b,Hossen2014c,Hossen2014d,Hosen2016}.
The existence of ions in astrophysical plasmas has been affirmed
experimentally by detecting of a noble gas molecular ion in the
crab nebula \cite{Barlow2013}. EI plasmas exists in Milky way
galaxy \cite{Burns1983}, polar region of neutron stars
\cite{Michel1991}, active galactic nuclei \cite{Miller1987},
pulsar magnetosphere \cite{Goldreich1969} and the early universe
\cite{Rees1983} etc. Ion acoustic waves play a vital role to
study the nonlinear features of localized electrostatic
disturbances in laboratory plasmas
\cite{Lonngren1983,Nakamura1985,Koga1993} as well as in space
plasmas \cite{Witt1983,Qian1988,Marchenko1995}.

Generally, the particle distribution near equilibrium in a plasma
system is often considered to be  Maxwellian, for modeling
purposes. But from the space plasma observations, it has been
proven that the presence of the ion and electron populations as
astrophysical plasma contexts are far off from their thermal
equilibrium state. Due to the effect of external forces or to wave
particle interaction in numerous space plasma observations
\cite{Vocks2003,Gloeckler2006} and laboratory experiments
\cite{Yagi1997,Preische1996} indicate the appearance of highly
energetic (superthermal) particles. The existence of accelerated,
energetic (superthermal) particles in the measurement of electron
distribution in near-Earth space environments suggest a (often
strong) deviation from Maxwellian equilibrium
\cite{Maksimovic1997,Gloeckler2006,Chaston1997}. So Maxwellian
Boltzmann distribution is possibly imperfect for explicating the
interaction of superthermal particles. However, the plasma system
containing higher energetic (Superthermal) particle with energies
greater than the energies of the particles exist in thermal
equilibrium can be fitted more appropriately via the $\kappa$
(kappa) type of Lorentzian distribution function (df)
\cite{Vasyliunas1968,Summers1991,Hellberg2009} than via the
thermal Maxwellian df where the real parameter $\kappa$ measures
the deviation from a Maxwellian distribution (the smaller the
value, the larger the deviation from a Maxwellian, in fact
attained for infinite $\kappa$). Hence we focus on a plasma
system with superthermal particles modelled by a $\kappa$-
distribution \cite{Hellberg2009}.

The isotropic three-dimensional (3D) kappa velocity distribution
of particles of mass m is of the form
\begin{eqnarray}
&&\hspace*{-5mm}F_k(v)=\frac{\Gamma(k+1)}{(\pi k \omega)^{(3/2)}
\Gamma (k-1/2)}(1+\frac{v^2}{k \omega^2})^{-(k+1)},
\end{eqnarray}

\noindent where, $F_k$ symbolizes the kappa distribution function,
$\Gamma$ is the gamma function, $\omega$ shows the most probable
speed of the energetic particles, given by $\omega =
[(2k-3/k)^{1/2} (k_B T/m)^{1/2} ]$, with $T$ being the
characteristic kinetic temperature and $\omega$ is related to the
thermal speed $V_t = (k_B T/m)^{1/2} $ and, the parameter $k$
represents the spectral index \cite{Cattaert2007} which defines
the strength of the superthermality. The range of this parameter
is $3/2 < k <
 \infty$ \cite{Alam2013}. In the limit $k \rightarrow \infty$ \cite{Basu2008,Baluku2012} the kappa distribution
function reduces to the well-known Maxwellian{ Boltzmann
distribution.

Recently, a numerous investigations have been made by many
authors on ion-acoustic solitary waves (IASWs) with
single-temperature superthermal (kappa distributed) electrons
\cite{Hussain2012,Shahmansouri2013,Sultana2011}. Schippers
\textit{et al.} \cite{Schippers2008} have combined a hot and a
cold electron component, while both electrons are kappa
distributed and found a best fit for the electron velocity
distribution. Baluku \textit{et al.} \cite{Baluku2011} used this
model as base of a kinetic theory study for electron-acoustic
waves in Saturn's magnetosphere and then they have studied IA
solitons in a plasma with two-temperature kappa distributed
electrons. Pakzad \cite{Pakzad2011} studied a dissipative plasma
system with superthermal electrons and positrons and, found the
effects of ion kinematic viscosity and the superthermal parameter
on the IA waves. Tasnim \textit{et al.}
\cite{Tasnim2013a,Tasnim2013b} also considered two-temperature
nonthermal ions and discussed the properties of shock waves.
Masud \textit{et al.} \cite{Masud2013} have studied the
characteristic of DIA shock waves in an unmagnetized dusty plasma
consisting of negatively charged static dust, inertial ions,
positively charged positrons following Maxwellian distribution,
and superthermal electrons with two distinct temperatures. Lu
\textit{et al.} \cite{Lu2005} studied the EA solitary structures
with positive charged potentials in a three-component plasma
system containing cold, hot, and beam electrons. By using the
small-k expansion perturbation method, the three-dimensional
stability of dust-ion acoustic solitary waves in a magnetized
multicomponent dusty plasma containing negative heavy ions and
stationary variable-charge dust particles was analyzed by
El-Taibany \textit{et al.} \cite{El-Taibany2011}. Shahmansouri
\cite{Shahmansouri2012} investigated the basic properties of IA
waves in an unmagnetized plasma including cool ions and hot ions
with kappa distributed electrons using small amplitude techniques
and observed that the suprathermality effects play a pivotal role
as the properties of IA solitons significantly affect through the
ion and the electron superthermal index. Since, dynamical heavy
particles and higher order nonlinearity play a vital role to
describe the different nonlinear structures of astrophysical
objects. Therefore, our main intention to carry the higher order
nonlinearity by deriving extention of mK-dV equation, namely
further mK-dV (fmK-dV) and also considering the dynamics of heavy
particles to describe HIASWs in such plasma system under
consideration. Recently, Ema \textit{et al.}
\cite{Ema2015a,Ema2015b} studied the effect of adiabacity on the
heavy ion acoustic (HIA) solitary and shock waves in a strongly
coupled nonextensive plasma. They observed that the roles of the
adiabatic positively charged heavy ions, and nonextensivity of
electrons have significantly modified the basic features (viz.,
polarity, amplitude, width, etc.) of the HIA solitary/shock
waves. Hossen \textit{et al.}
\cite{Hossen2014m,Hossen2014n,Hossen2015o} considered positively
charged static heavy ions in a relativistic degenerate plasma and
rigorously investigated the basic features of solitary and shock
structures. By considering superthermal electrons and adiabatic
heavy ions Shah \textit{et al.} \cite{Shah2015a,Shah2016}
investigated the basic features of HIA solitary and shock waves
by considering both planar and nonplanar geometry.

Up to the best of our knowledge, no theoretical investigations
have been worked out to study of HIASWs with two temperature
superthermal (kappa distributed) electrons in EI plasma.
Therefore, in our present work, we attempt to study the basic
features of HIA waves by deriving the magnetized Korteweg-de
Vries (K-dV) and magnetized modified K-dV (mK-dV) equations in EI
plasma containing superthermal electrons and Heavy ion.
%%%%%%%%%%%%%%%%%%%%%%%%%%%%%%%%%%%%%%%%%%%%%%%%%%%%%%%%%%%%%%%%%%
\section{Theoretical model and basic equations}

We consider a three component magnetized plasma system containing
positively charged heavy ions and kappa distributed electrons with
two distinct temperatures. Therefore, at equilibrium condition,
$n_{hi0}=n_{ce0}+n_{he0}$, where $n_{hi0}$ is the unperturbed
heavy ion number density, $n_{ce0}$ and $n_{he0}$ are the density
of unperturbed lower and higher temperature electron
respectively. The dynamics of the heavy ion acoustic waves are
described by the following (normalized) equations:
\begin{eqnarray}
&&\hspace*{-5mm}\frac{\partial n_h}{\partial t} + \nabla
.({n_h}{\textbf{u}_h})= 0,\label{A1a}\\
&&\hspace*{-5mm}\frac{\partial \textbf{u}_h}{\partial t} +
({\textbf{u}_h}.\nabla)u_h=- \nabla \phi+
\alpha({\textbf{u}_h}\times{\hat{z}}).\label{A1b}
\end{eqnarray}
The Poisson's equation can be written as

\begin{eqnarray}
&&\hspace*{-5mm}\nabla^2\phi=\mu_{ce}\left(1-\frac{\sigma_1\phi}{k_{ce}-\frac{3}{2}}\right)^{-k_{ce}
+\frac{1}{2}}-n_h\nonumber\\&&\hspace*{-5mm}
+\mu_{he}\left(1-\frac{\sigma_2\phi}{k_{he}-\frac{3}{2}}\right)^{-k_{he}+\frac{1}{2}},\label{A1c}
\end{eqnarray}

\noindent where, $n_h$ is the heavy ion particle number density
normalized by its equilibrium value $n_{hi0}$, $\textbf u_h$ is
the heavy ion fluid speed normalized by
$C_h=(k_BT_{ef}/m_h)^{1/2}$, $\phi$ is the electrostatic wave
potential normalized by $k_BT_{ef}/e, \sigma_1=T_{ef}/T_{ce},
\sigma_2=T_{ef}/T_{he}, \mu_{ce}=n_{ce0}/n_{hi0},
\mu_{he}=n_{he0}/n_{hi0}, \alpha =\omega_{hc}/\omega_{ph} $, where
$T_{ef}=n_{e0}T_{ce}T_{he}/(n_{ce0}T_{he}+n_{he0}T_{ce})$. It
should be noted that $n_{e0}$ is the total electron number
density at equilibrium and $T_{ce}(T_{he})$ is the lower (higher)
electron temperature, $T_{ef}$ is the effective temperature of two
electrons, $k_{ce}$ and $k_{he}$ are spectral index,
$\omega_{hc}$ is the heavy ion cyclotron frequency, $\omega_p$ is
the plasma frequency, $k_B$ is the Boltzmann constant and e is
the magnitude of the electron charge. The time variable ($t$) is
normalized by ${\omega_{ph}}=(4 \pi n_{hi0}e^2/m_h)^{1/2}$, and
the space variable ($x$) is normalized by $\lambda_{h}=(m_ec^2/4
\pi n_{hi0}e^2)^{1/2}$.

%%%%%%%%%%%%%%%%%%%%%%%%%%%%%%%%%%%%%%%%%%%%%%%%%%%%%%%%%%%%%%%%%%%%%%%%%%%%%%%%%%%%%%%%
\section{Nonlinear Equations}
\subsection{Derivation of the Magnetized K-dV Equation}
At first we use reductive perturbation method to derive the well
known K-dV equation. We apply the reductive perturbation
technique in which independent variables are stretched as

\begin{eqnarray}
&&\xi=\epsilon^{1/2}(L_x\hat{x}+L_y\hat y+L_z\hat z -V_pt),
\label{A1d}\\
&&\tau={\epsilon}^{3/2}t, \label{A1e}
\end{eqnarray}

\noindent where, $V_p$ is the phase velocity of HIA wave and
$\epsilon$ is a smallness parameter measuring the weakness of the
dispersion $(0<\epsilon<1)$ and $L_x$, $L_y$, and $L_z$ are the
directional cosines of the wave vector k along the x, y, and z
axes respectively, so that $L_x^2+L_y^2+L_z^2= 1$. For a dynamical
equation, we also expand the perturbed quantities $n_h$, $u_h$
and $\phi$ in power series of $\epsilon$. We may expand $n_h$,
$u_h$, and $\phi$ in power series of $\epsilon$ as

\begin{eqnarray}
&&n_h=1+\epsilon n_h^{(1)}+\epsilon^{2}n_h^{(2)}+ \cdot \cdot
\cdot, \label{A1f}\\
&&u_{hx,y}=0+\epsilon^{3/2}
u_{hx,y}^{(1)}+\epsilon^{2}u_{hx,y}^{(2)}+\cdot \cdot \cdot,
\label{A1g}\\
&&u_{hz}=0+\epsilon u_{hz}^{(1)}+\epsilon^{2}u_{hz}^{(2)}+\cdot
\cdot \cdot,
\label{A1h}\\
&&\phi=0+\epsilon\phi^{(1)}+\epsilon^{2}\phi^{(2)}+\cdot \cdot
\cdot    . \label{A1i}
\end{eqnarray}

 Now, applying Eqs. (\ref{A1d})-(\ref{A1i}) into Eqs. (\ref{A1a}) -
(\ref{A1c}) and taking the lowest order coefficient of $\epsilon$,
we obtain, $u_{hz}^{(1)}={L_z \phi^{(1)}}/{V_p}$,
$n_{h}^{(1)}={L_z^2 \phi^{(1)}}/{V_p^2}$ and
$V_p=L_z/\sqrt{\mu_{ce}c_1+\mu_{he}d_1}$ represents the dispersion
relation for the HIA waves that move along the propagation vector
$k$. To the next higher order of $\epsilon$, we obtain a set of
equations after using the value of $u_{hz}^{(1)}$,
$n_{hz}^{(1)}$, $V_p$ and taking the z-component of momentum
equation as

\begin{eqnarray}
&&\frac{\partial n_h^{(1)}}{\partial \tau}-v_p \frac{\partial
n_h^{(2)}}{\partial \xi}+L_x\frac{\partial u_{hx}^{(2)}}{\partial
\xi}\nonumber\\&&+L_y\frac{\partial u_{hy}^{(2)}}{\partial
\xi}+L_z\frac{\partial u_{hz}^{(2)}}{\partial
\xi}+L_z\frac{\partial {(n_h^{(1)}u_{hz}^{(1)})}}{\partial \xi}=0,
\label{A1j}\\
&&\frac{\partial u_{hz}^{(1)}}{\partial \tau}-v_p\frac{\partial
u_{hz}^{(2)}}{\partial \xi}+L_z u_{hz}^{(1)}\frac{\partial
u_{hz}^{(1)}}{\partial \xi}+L_z\frac{\partial
{\phi^{(2)}}}{\partial \xi}=0,
 \label{A1k}\\
&&\frac{\partial^2 \phi^{(1)}}{\partial
\xi^2}=\mu_{ce}c_1\phi^{(2)}+\mu_{ce}c_2\phi{^{(1)}}^2\nonumber\\&&-n_h^{(2)}+\mu_{he}d_1\phi^{(2)}+\mu_{he}d_2\phi{^{(1)}}^2.
\label{A1k}\
\end{eqnarray}
Again taking the co-efficient of $\epsilon^2$ for x-and y-
component from momentum equation we get
\begin{eqnarray}
 &&u_{hy}^{(2)}=\frac{L_yV_P}{\alpha^2}\frac{\partial^2\phi^{(1)}}{\partial\xi^2},
 \label{A1l}\\
 &&u_{hx}^{(2)}=\frac{L_xV_P}{\alpha^2}\frac{\partial^2\phi^{(1)}}{\partial\xi^2}.
\label{A1m}
\end{eqnarray}
Now combining the equation (\ref{A1j}) - (\ref{A1m}), we have a
equation of the form

\begin{eqnarray}
&&\frac{\partial\phi^{(1)}}{\partial \tau} + \lambda \phi^{(1)}
\frac{\partial \phi^{(1)}}{\partial \xi}+ \beta \frac{\partial^3
\phi^{(1)}}{\partial \xi^3}=0. \label{A1n}
\end{eqnarray}

This is well-known K-dV equation that describes the obliquely
propagating HIA waves in a magnetized plasma. Where
\begin{eqnarray}
&&\lambda=\frac{3L_z^2}{2V_p}-V_p,\label{A3m}\\
&&\beta=\frac{V_p^3}{2L_z^2}\left[1+\frac{(1-L_z^2)}{\alpha^2}\right].
\label{A10}\
\end{eqnarray}

Equation (15) which is known as K-dV (Korteweg-de Vries) equation
and the stationary localized solution of Eq. (15) is given by
\begin{eqnarray}
{\rm \phi^{(1)}}=\rm \phi_m{\rm[sech^{2}}(\frac{\xi}{\Delta})],
\label{solK-dV}
\end{eqnarray}

\noindent where the amplitude, $\phi_m=3u_{0}/\lambda$, and the
width, $\Delta=(4\beta/u_{0})^{1/2}$

\begin{eqnarray}
&&c_1=\frac{(2k_{ce}-1)\sigma_1}{2k_{ce}-3},\label{A1p}\\
&&d_1=\frac{(2k_{he}-1)\sigma_2}{2k_{he}-3},\label{A1q}\\
&&c_2=\frac{(2k_{ce}-1)(2k_{ce}+1)\sigma_1^2}{2(k_{ce}-3)^2},\label{A1r}\\
&&d_2=\frac{(2k_{he}-1)(2k_{he}+1)\sigma_2^2}{2(k_{he}-3)^2}.\label{A1s}
\end{eqnarray}

\subsection{Derivation of the Magnetized mK-dV Equation}

Same stretched co-ordinates is applied to obtain the mK-dV
equation as we used in K-dV equation in section IV (i.e., Eqs.(5)
and (6)) and also used the depended variables which is expanded as

\begin{eqnarray}
&&n_h=1+\epsilon^{1/2} n_h^{(1)}+\epsilon n_h^{(2)}+\epsilon^{3/2}
n_h^{(3)} +\cdot \cdot
\cdot, \label{A1t}\\
&&u_{hx,y}=0+\epsilon
u_{hx,y}^{(1)}+\epsilon^{3/2}u_{hx,y}^{(2)}+\epsilon^{2}u_{hx,y}^{(3)}+\cdot
\cdot \cdot,
\label{A1u}\\
&&u_{hz}=0+\epsilon^{1/2} u_{hz}^{(1)}+\epsilon
u_{hz}^{(2)}+\epsilon^{3/2} u_{hz}^{(3)}+\cdot \cdot \cdot,
\label{A1v}\\
&&\phi=0+\epsilon^{1/2}\phi^{(1)}+\epsilon\phi^{(2)}+\epsilon^{3/2}\phi^{(3)}\cdot
\cdot \cdot . \label{A1w}
\end{eqnarray}
We find the same expressions of $n_{h}^{(1)}$, $u_{hz}^{(1)}$,
$u_{hx,y}^{(1)}$, $u_{hx,y}^{(2)}$ and $V_p$ as like as that of
K-dV equation. To the next higher order of $\epsilon$, by using
the values of $n_{h}^{(1)}$, $u_{hz}^{(1)}$, $u_{hx,y}^{(1)}$,
$u_{hx,y}^{(2)}$ and $V_p$ we obtain a set of equations which can
be simplified as
\begin{eqnarray}
&&\hspace*{-38mm}u_{hz}^{(2)}=\frac{L_z^3{\phi^{(1)}}^2}{2{V_{p}}^3}+\frac{L_z
{\phi^{(2)}} }{V_p},\label{A1x}\\
&&\hspace*{-38mm} n_h^{(2)}=\frac{3L_z^4
{\phi^{(1)}}^2}{2V_p^4}+\frac{L_z^2 {\phi^{(2)}}}{V_p^2},\label{A1y}\\
\rho^{(2)}=-\frac{1}{2}A{\phi^{(1)}}^2=0,\label{A1z}\
\end{eqnarray}
where,
$$A=\left[\frac{3L_z^4}{2V_p^4}-(\mu_{ce}c_2+\mu_{he}d_2)\right].$$
To the next higher order of $\epsilon$, we obtain a set of
equations

\begin{eqnarray}
&&\hspace*{-8mm}\frac{\partial n_h^{(1)}}{\partial
\tau}-V_p\frac{\partial n_h^{(3)}}{\partial\xi}+L_x\frac{\partial
u_{hx}^{(2)}}{\partial\xi}+L_x\frac{\partial}{\partial\xi}(n_h^{(1)}u_{hx}^{(1)})\nonumber\\
&&\hspace*{-8mm}+L_y\frac{\partial
u_{hy}^{(2)}}{\partial\xi}+L_y\frac{\partial}{\partial\xi}(n_h^{(1)}u_{hy}^{(1)})+L_z\frac{\partial
u_{hz}^{(3)}}{\partial\xi}\nonumber\\
&&\hspace*{-8mm}+L_z\frac{\partial}{\partial\xi}(n_h^{(1)}u_{hz}^{(2)})+L_z\frac{\partial}{\partial\xi}(n_h^{(2)}u_{hz}^{(1)})=0,
\label{A2a}\\
&&\hspace*{-8mm}\frac{\partial u_{hz}^{(1)}}{\partial
\tau}-V_p\frac{\partial
u_{hz}^{(3)}}{\partial\xi}+L_z\frac{\partial}{\partial\xi}(u_{hz}^{(1)}u_{hz}^{(2)})+L_z\frac{\partial\phi^{(3)}}{\partial\xi}=0,
\label{A2b}\\
&&\hspace*{-8mm}\frac{\partial^2\phi^{(1)}}{\partial\xi^2}=(\mu_{ce}c_1+\mu_{he}d_1)\phi_3+2(\mu_{ce}c_2 \nonumber\\
&&\hspace*{-8mm}+\mu_{he}d_2)\phi^{(1)} \phi^{(2)} -n_h^{(3)}.
\label{A2c}
\end{eqnarray}

\begin{figure}[t!]
\centerline{\includegraphics[width=6.8cm]{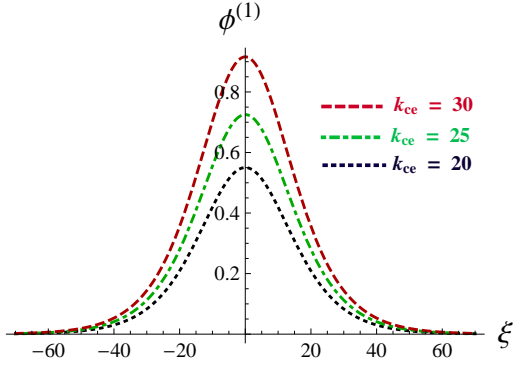}} \caption{(Color
online) Showing the variation of the positive potential of
magnetized K-dV solitons $\phi^{(1)}$ with $k_{ce}$ for
$\mu_{ce}>\mu_{c}$, $u_0=0.01$, $\sigma_1 = 2.5$, $\sigma_2=0.1$,
$\mu_{he}=0.04$, $k_{he}= 2$, $\delta=5$ and $\alpha =0.5$ .}
\label{A2d}
\end{figure}

\begin{figure}[t!]
\centerline{\includegraphics[width=6.8cm]{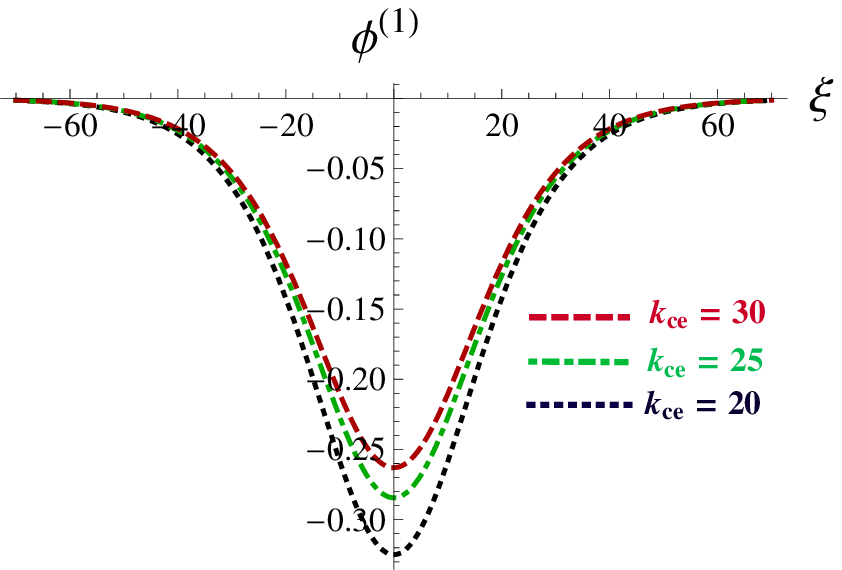}} \caption{(Color
online) Showing the variation of the negative potential of
magnetized K-dV solitons $\phi^{(1)}$ with $k_{ce}$ for
$\mu_{ce}<\mu_{c}$. The other plasma parameters are kept fixed.}
\label{A2e}
\end{figure}

\begin{figure}[t!]
\centerline{\includegraphics[width=6.8cm]{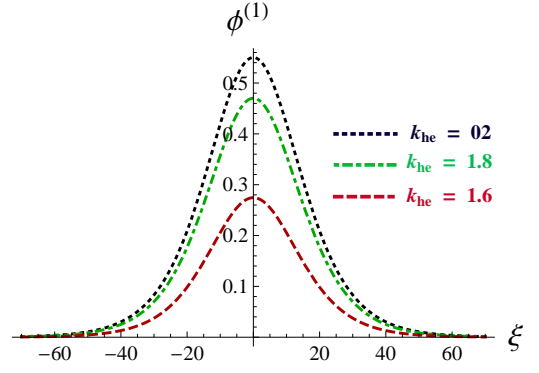}} \caption{(Color
online) Showing the variation of the positive potential of
magnetized K-dV solitons $\phi^{(1)}$ with $k_{he}$ for
$\mu_{ce}>\mu_{c}$ and $k_{ce}=20$. The other plasma parameters
are kept fixed.} \label{A2f}
\end{figure}

\begin{figure}[t!]
\centerline{\includegraphics[width=6.8cm]{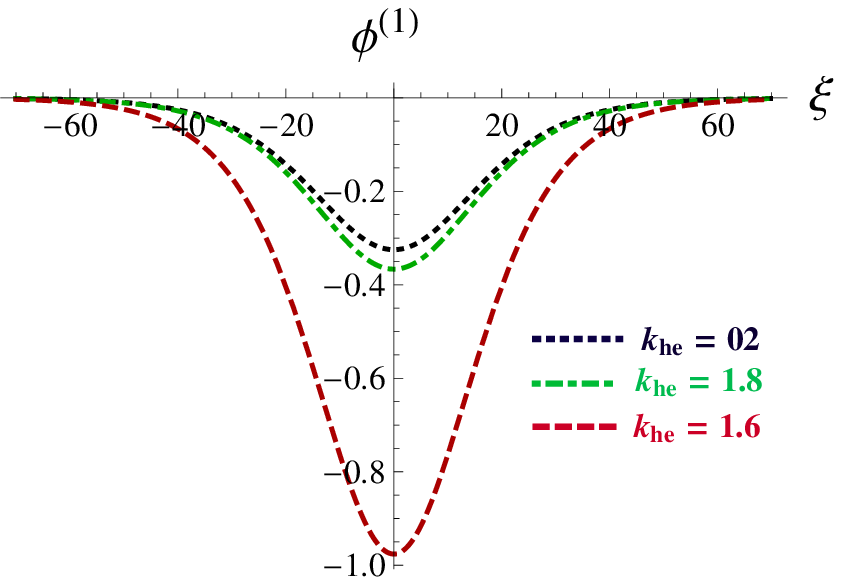}} \caption{(Color
online) Showing the variation of negative potential of magnetized
K-dV solitons $\phi^{(1)}$ with $k_{he}$ for $\mu_{ce}<\mu_{c}$.
The other plasma parameters are kept fixed.} \label{A2g}
\end{figure}

Now, combining Eqs. (\ref{A2a}) - (\ref{A2c}) we obtain a equation
of the form

\begin{eqnarray}
&&\hspace*{-10mm}\frac{\partial\phi^{(1)}}{\partial \tau} +
\alpha_1 \alpha_3\phi^{(1)2} \frac{\partial \phi^{(1)}}{\partial
\xi}+\alpha_2\alpha_3\frac{\partial^3 \phi^{(1)}}{\partial
\xi^3}=0. \label{DIAK-dV}
\end{eqnarray}

This is well-known mK-dV equation that describes the obliquely
propagating HIA waves in a magnetized plasma. where $\alpha_1$,
$\alpha_2$ and $\alpha_3$ are given by

\begin{eqnarray}
&&\alpha_1=\frac{15L_z^6}{2V_p^6},
\label{A2h}\\
&&\alpha_2=1+\frac{(1-L_z^2)}{\alpha^2},
\label{A2i}\\
&&\alpha_3=\frac{V_p^3}{2L_z^2}.
\end{eqnarray}

\begin{figure}[t!]
\centerline{\includegraphics[width=6.8cm]{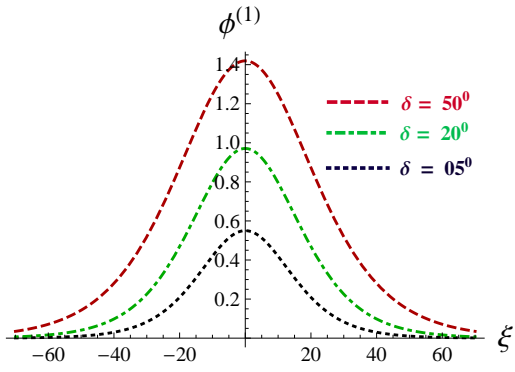}} \caption{(Color
online) Showing the variation of the positive potential of
magnetized K-dV solitons $\phi^{(1)}$ with obliqueness $\delta$.
The other plasma parameters are kept fixed.} \label{A2j}
\end{figure}

\begin{figure}[t!]
\centerline{\includegraphics[width=6.8cm]{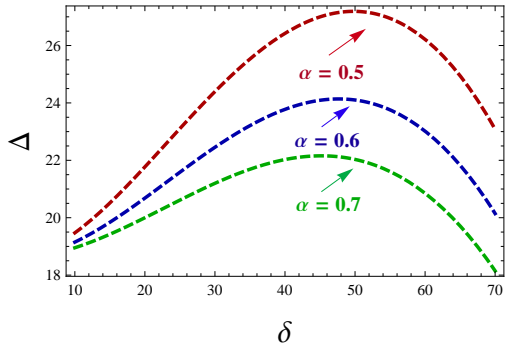}} \caption{(Color
online) Showing the variation of the width of magnetized K-dV
solitons $\phi^{(1)}$ with $\alpha$. The other plasma parameters
are kept fixed.} \label{A2k}
\end{figure}

\begin{figure}[t!]
\centerline{\includegraphics[width=6.8cm]{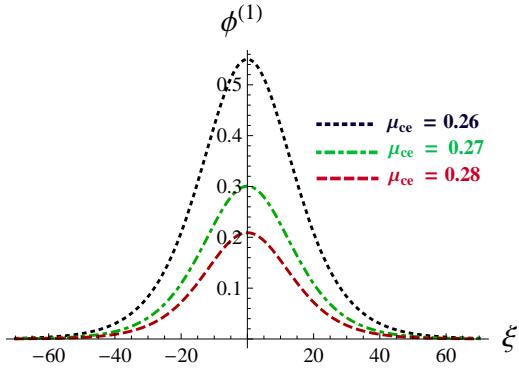}} \caption{(Color
online) Showing the variation of the amplitude of magnetized K-dV
solitons with different values of $\mu_{ce}$. The other plasma
parameters are kept fixed.} \label{A2l}
\end{figure}

\begin{figure}[t!]
\centerline{\includegraphics[width=6.8cm]{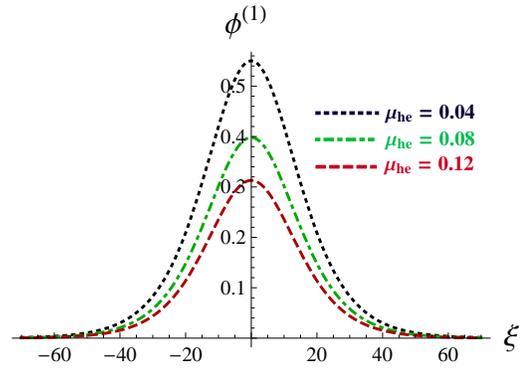}} \caption{(Color
online) Showing the variation of positive potential K-dV solitons
with different values of $\mu_{he}$. The other plasma parameters
are kept fixed.} \label{A2m}
\end{figure}

\begin{figure}[t!]
\centerline{\includegraphics[width=6.8cm]{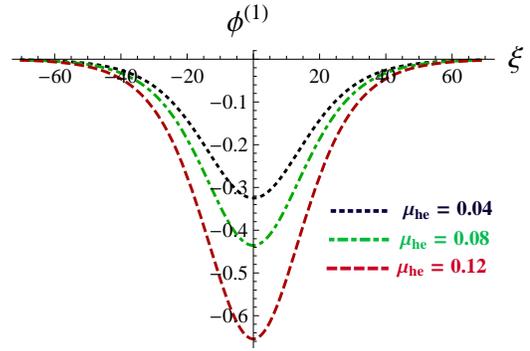}} \caption{(Color
online) Showing the variation of negative potential K-dV solitons
with different values of $\mu_{he}$. The other plasma parameters
are kept fixed except.} \label{A2n}
\end{figure}

The stationary solitary wave solution of standard mK-dV equation
is obtained by considering a frame $\xi=\eta-u_{0}T$ (moving with
speed $u_{0}$) and the solution is,
\begin{eqnarray}
{\rm \phi^{(1)}}=\rm \phi_m{\rm[sech}(\frac{\xi}{\varpi})],
\label{solK-dV}
\end{eqnarray}

\noindent where the amplitude, $\phi_m=\sqrt{(6u_{0}/\alpha_1
\alpha_3)}$ and the width $\varpi= \phi_m {\sqrt{(\alpha_1/6)}}$.

\begin{figure}[t!]
\centerline{\includegraphics[width=6.8cm]{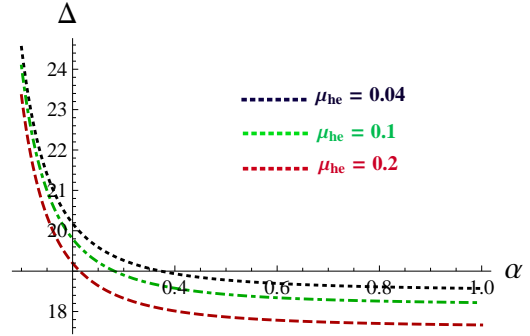}} \caption{(Color
online) Showing the variation of width (delta) of magnetized K-dV
solitons with $\mu_{he}$ . The other plasma parameters are kept
fixed.} \label{A2o}
\end{figure}

\begin{figure}[t!]
\centerline{\includegraphics[width=6.8cm]{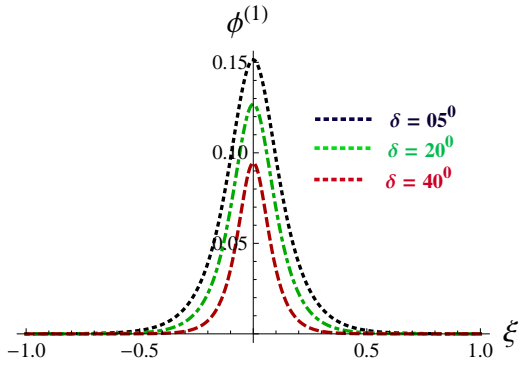}} \caption{(Color
online) Showing the variation of the amplitude of magnetized
mK-dV solitons with with obliqueness $\delta$. The other plasma
parameters are kept fixed.} \label{A2p}
\end{figure}

\begin{figure}[t!]
\centerline{\includegraphics[width=6.8cm]{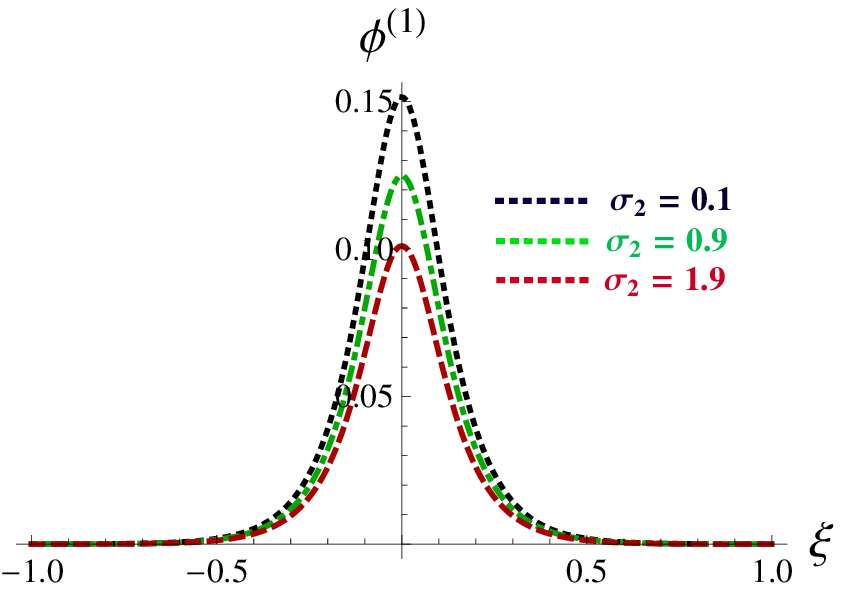}} \caption{(Color
online) Showing the variation of the amplitude of magnetized
mK-dV solitons with $\sigma_2$ keeping other plasma parameters
fixed.} \label{A2q}
\end{figure}

\begin{figure}[t!]
\centerline{\includegraphics[width=6.8cm]{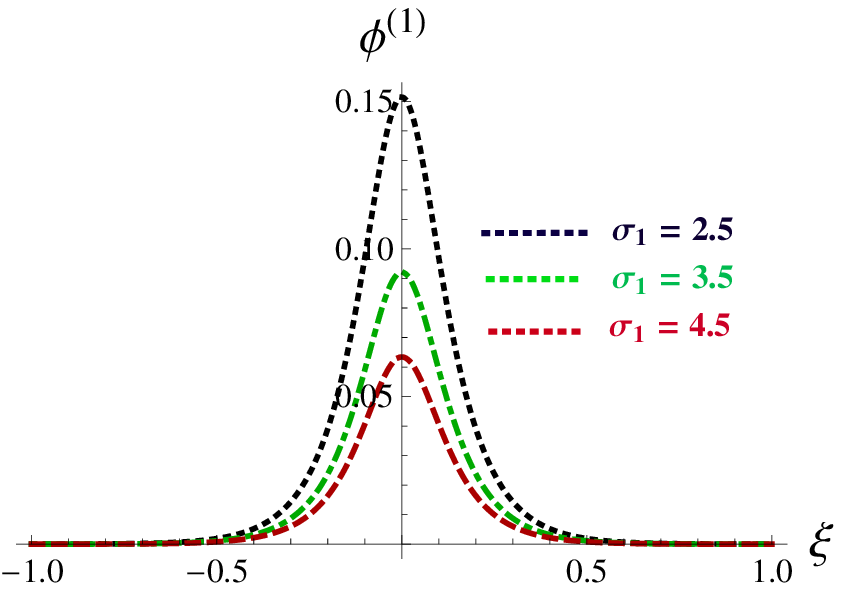}} \caption{(Color
online) Showing the variation of the amplitude of magnetized
mK-dV solitons with $\sigma_1$ keeping other plasma parameters
fixed.} \label{A2r}
\end{figure}

\begin{figure}[t!]
\centerline{\includegraphics[width=6.8cm]{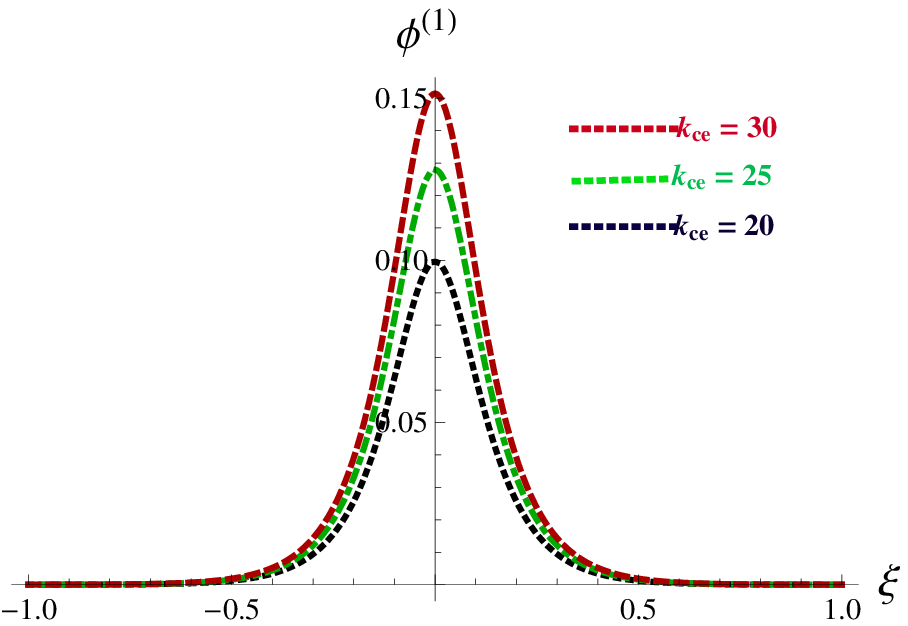}} \caption{(Color
online) Showing the variation of the amplitude of magnetized
mK-dV solitons with $k_{ce}$ keeping other plasma parameters
fixed.} \label{A2s}
\end{figure}

\begin{figure}[t!]
\centerline{\includegraphics[width=6.8cm]{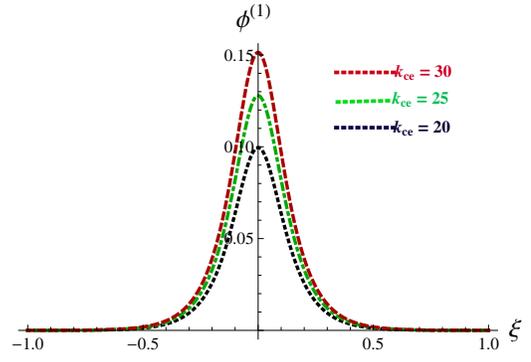}} \caption{(Color
online) Showing the variation of the amplitude of magnetized
fmK-dV compressive solitons with $k_{ce}$ keeping other plasma
parameters fixed.} \label{15}
\end{figure}

\subsection{Derivation of the Magnetized further mK-dV Equation}
In order to trace the vicinity of the critical heavy ion density,
neither the magnetized KdV nor magnetized mKdV equation is
suitable for describing the HIASWs. To examine the propagation of
HIASWs in the vicinity of critical heavy ion density region, we
assume a nonzero value for the ${\phi^{(1)}}^2$ of the Poisson's
equation and finally one can derive the following nonlinear
equation

\begin{eqnarray}
&&\hspace*{-1mm}\frac{\partial\phi^{(1)}}{\partial \tau}
+[\lambda \alpha_3 \phi^{(1)}+\alpha_1 {\alpha_3}^2
\{\phi^{(1)}\}^2] \frac{\partial \phi^{(1)}}{\partial \xi} \nonumber\\
&&\hspace*{18mm} + \alpha_3 \frac{\partial^3 \phi^{(1)}}{\partial
\xi^3}= 0. \label{fK-dV}
\end{eqnarray}

Eq. (39) is known as magnetized further mK-dV (fmK-dV) equation
and the values of the constants are already mentioned in the
previous. The compressive soliton solution
(\cite{El-Shewy2011,Shah2015}) of Eq. (39) is

\begin{eqnarray}
{\rm \phi^{(1)}_+}=\frac{6u_0/\lambda \alpha_1 \alpha_3}
{\phi^{(1)}sinh^2(\sqrt{u_0/2\lambda}\xi)-\phi^{(2)}cosh^2(\sqrt{u_0/2\lambda}\xi)},
\label{solfmK-dV1}
\end{eqnarray}
and the rarefactive soliton solution
(\cite{El-Shewy2011,Shah2015})

\begin{eqnarray}
{\rm \phi^{(1)}_-}=\frac{6u_0/\lambda \alpha_1
\alpha_3}{\phi^{(2)}sinh^2(\sqrt{u_0/2\lambda}\xi)-\phi^{(1)}cosh^2(\sqrt{u_0/2\lambda}\xi)}.
\label{solfmK-dV2}
\end{eqnarray}

%%%%%%%%%%%%%%%%%%%%%%%%%%%%%%%%%%%%%%%%%%%%%%%%%%%%%%%%%%%%%%%%%%%%%%%%%%%%%.
\section{Discussion and Results}

We have considered a magnetized plasma system consisting of
inertial heavy ions and  kappa distributed electrons of two
distinct  temperatures. We have derived  the magnetized K-dV,
mK-dV and fmK-dV-type partial differential equations by using the
reductive perturbation method to investigate  the basic features
(i.e., amplitude, width and polarity etc.) of such a plasma
system. The magnetized K-dV, mK-dV and fmK-dV equations are solved
to set out the fascinating features of HIASWs. Then these
solutions are analyzed by taking the effect of different plasma
parameters like relative temperature-ratio of electrons (i.e.,
$\sigma_1$ and $\sigma_2$ ), superthermal parameter $k$ (i.e.,
$k_{ce}$ and $k_{he}$), relative electron number density ratio to
heavy ion $\mu$ (i.e., $\mu_{ce}$ and $\mu_{he}$ ) etc. The
results, which have been obtained from this theoretical
investigation, can be pin-pointed as follows:

\begin{enumerate}

\item{The magnetized K-dV and fmK-dV equations support
HIA solitary waves with either  compressive (positive) or
rarefactive (negative) which depends on the critical value
$\mu_c$, but for magnetized mK-dV equation only  compressive SWs
are formed.}

\item{We have numerically obtained here that for $\lambda=0$, the
amplitude of the K-dV solitons become infinitely large, and the
K-dV solution is no longer valid at $\lambda\simeq0$. In our
present investigation, we have found that for
$\mu_c=\mu_{ce}=0.24$, the amplitude of the SWs breaks down due
to the vanishing of the nonlinear coefficient $\lambda$. We have
observed that at $\mu_c>0.24$, positive (compressive) potential
SWs exist, whereas at $\mu_c<0.24$, negative (rarefactive) SWs
exist (shown in Figs. 1-9).}

\item{It is observed that, the amplitude and width of both
positive and negative potential magnetized K-dV SWs are totally
depend on superthermal parameter of cold electron $k_{ce}$ and
hot electron $k_{he}$. It is found that the amplitude and width
of  the  positive and negative potential magnetized K-dV SWs
increases with the increasing values of $k_{ce}$, as well as
increasing the values of $k_{he}$ (shown in Figs. 1-4).}

\item{It is noticed that the amplitude and width of the positive
potential  of the magnetized K-dV soliton increases with the
increase in obliqueness of the wave propagation $\delta$ (see
Fig. 5).}

\item{It is observed that as the value of $\delta$ increases, the
amplitude of the solitary waves increases, while their width
increases for the lower range of $\delta$ (from $0^{\circ}$ to
about $50^{\circ}$), and decreases for its higher range (from
$50^{\circ}$ to about $90^{\circ}$). As $\delta
\rightarrow90^{\circ}$, the width goes to $0$, and the amplitude
goes to $\infty$. It is likely that for large angles, the
assumption that the waves are electrostatic is no longer valid,
and we should look for fully electromagnetic structures. Our
present investigation is only valid for small value of $\delta$
but invalid for arbitrary large value of $\delta$. In case of
larger values of $\delta$, the wave amplitude becomes large
enough to break the validity of the reductive perturbation
method.}

\item{From the Fig. 6, it is seen that the amplitude and width of the
solitary profile decreases with increasing values of number
density ratio of cold electron to heavy ion $\mu_{ce}$.}

\item{It is found from Figs. 7 and 8, that the amplitude and width of both
positive and negative potential magnetized K-dV SWs depend on the
values of number density ratio  of hot electron to heavy ion
$\mu_{he}$. It is seen that the amplitude and width of  the
positive and negative potential magnetized K-dV SWs increases with
the decreasing values of $\mu_{he}$.}

\item{Variation of the width of magnetized K-dV SWs for different values of $\mu_{he}$
with $\alpha$ is depicted in fig. 10. It is obtained that the
width of magnetized K-dV SWs decreases with the increasing values
of $\mu_{he}$.}

\item{In our present investigation, we have observed only positive potential magnetized
mK-dV SWs. It is found that the amplitude and width of magnetized
mK-dV SWs decreases with the increasing value of obliqueness
$\delta$ (see Fig.11).}

\item{We have analyzed and found that the basic features of HIASWs depend on the relative
temperature ratio of electrons(i.e., $\sigma_1$ and $\sigma_2 $).
It is found that the amplitude and width of magnetized mK-dV SWs
decreases with the increasing value of $\sigma_1$ and $\sigma_2$
(see Figs. 12 and 13).}

\item{The variation of amplitude and width of magnetized mK-dV SWs which depends on
superthermal parameter of cold electron $k_{ce}$ is depicted in
Fig. 14. It is observed that the amplitude and width of magnetized
mK-dV SWs decreases with the increasing value of $\sigma_1$ and
$\sigma_2$.}

\item{The comparison among the K-dV, mK-dV and fmK-dV solitary amplitude is also important. It is observed that
the amplitude of solitary waves decreases with the increasing of
nonlinearities (see Figs. 1, 14, and 15).}
\end{enumerate}

In compressed, HIASWs have been studied in a magnetized space
plasma. Since astrophysical objects are consists of heavy ions,
the results of our present assessment can be applicable to the
investigation on the HIASWs in such magnetized astrophysical
objects including neutron stars, pulsar magnetosphere
\cite{Kundu2011}, Saturn's magnetosphere \cite{Baluku2012},
peculiar velocities of galaxy clusters etc. where the effect of
two temperature superthermal electrons play a crucial role.

\section{Acknowledgments}

M. Sarker, B. Hosen and M. R. Hossen are profoundly grateful to
the Ministry of Science and technology (Bangladesh) for awarding
the National Science and Technology (NST) fellowship.

%%%%%%%%%%%%%%%%%%%%%%%%%%%%%%%%%%%%%%%%%%%%%%%%%%%%%%%%%%%%%%%%%%%%%%%%%%%%%%%%%%%%%%%%%%%%%%%%%%%%%%%%%%%%%%%%%%%%%%%

\end{document}